\documentclass[preprint,onecolumn,aps,pre,eqsecnum,longbibliography,showpacs,showkeys,a4paper]{revtex4-1}  

\usepackage{hyperref} 
\usepackage{amsmath}
\usepackage{amssymb}
\usepackage{graphicx}

\renewcommand{\d}{{\,\rm d}}
\newcommand{\e}{{\rm e}}
\newcommand{\p}{{\partial}}
\newcommand{\VEC}[1]{{\mathbf{#1}}}

\newcommand{\meanx}[1]{\overline{#1}}
\newcommand{\mpbracket}{{\ensuremath{\genfrac{}{}{0pt}{1}{-}{\scriptstyle(\kern-1pt+\kern-1pt)}}}}
\newcommand{\pmbracket}{{\ensuremath{\genfrac{}{}{0pt}{1}{+}{\scriptstyle(\kern-1pt-\kern-1pt)}}}}

\DeclareGraphicsExtensions{.png,.pdf,.eps}
\graphicspath{{pictures/}}

\begin{document}

\title{An explanation of interference effects in the double slit experiment: Classical trajectories plus ballistic diffusion caused by zero-point fluctuations}

\author{Gerhard \surname{Gr\"ossing}}
\email[E-mail: ]{ains@chello.at}
\homepage[Visit: ]{http://www.nonlinearstudies.at/}
\author{Siegfried \surname{Fussy}}
\email[E-mail: ]{ains@chello.at}
\homepage[Visit: ]{http://www.nonlinearstudies.at/}
\author{Johannes \surname{Mesa Pascasio}}
\email[E-mail: ]{ains@chello.at}
\homepage[Visit: ]{http://www.nonlinearstudies.at/}
\author{Herbert \surname{Schwabl}}
\email[E-mail: ]{ains@chello.at}
\homepage[Visit: ]{http://www.nonlinearstudies.at/}
\affiliation{%
Austrian Institute for Nonlinear Studies\\
Akademiehof\\
Friedrichstr.~10, 1010 Vienna, Austria
\vspace*{5.0cm}
}

\date{\today}

\begin{abstract}
A classical explanation of interference effects in the double slit experiment is proposed.
We claim that for every single ``particle'' a thermal context can be defined, which reflects its embedding within boundary conditions as given by the totality of arrangements in an experimental apparatus.
To account for this context, we introduce a ``path excitation field'', which derives from the thermodynamics of the zero-point vacuum and which represents all possible paths a ``particle'' can take via thermal path fluctuations. 
The intensity distribution on a screen behind a double slit is calculated, as well as the corresponding trajectories and the probability density current. 
The trajectories are shown to obey a ``no crossing'' rule with respect to the central line, i.e., between the two slits and orthogonal to their connecting line. 
This agrees with the Bohmian interpretation, but appears here without the necessity of invoking the quantum potential.
\end{abstract}

                 
\maketitle

\renewcommand{\thesection}{\arabic{section}} 
\renewcommand{\thesubsection}{\arabic{section}.\arabic{subsection}}

\section{Introduction}
\label{sec:intro} 

Unless one assumed a purely pragmatic attitude and considered the formalism of quantum theory only as a set of rules to account for the empirical data, the challenge since the early days has been about developing a physical theory with a clear statement on \textit{what it is} that is moving, and under which relevant \textit{boundary conditions} and \textit{forces}. 

Actually inherited from debates about the nature of light in still earlier centuries, this question was also in the twentieth century largely characterized by whether quantum systems were, at the most fundamental level, particles or waves. A notable exception to these monistic approaches is the de~Broglie-Bohm interpretation of the quantum mechanical formalism  with its model of particle \textit{plus} ``guiding'' wave, i.e., a dualistic approach, albeit with an inherent asymmetry: the wave field acts on the particle, but not \textit{vice versa}, a feature unknown to any other part of physics.

Still, the modelling, or even the ontology, of all these approaches implies the existence of fundamental entities, be they particles or waves or something more complicated like, e.g., ``wavicles'', or ``beables'', etc..  
These entities are assumed to characterize physical processes over a vast range of orders of magnitude, including the yet unknown domains of very small space and time scales down to the Planck length (where, eventually, some new physics may be required). However, these are unwarranted assumptions only.

On the contrary, it may well be that there is practically an (almost?) infinite range of orders of magnitude, compared to the scales of today's highest resolutions. 
Facing such a scenario, or similar ones, it would be highly surprising, and even implausible, that at this point in history we should already know the ``most fundamental level'' of the laws of nature, i.e., in the form of the quantum physical laws.
Note, for example, that if standard quantum field theory were valid up to the Planck scale, one would obtain the ``famous'' discrepancy of 120 orders of magnitude between theoretical and experimental values of the vacuum energy~\cite{Carroll.2001}.
It would rather be much more plausible, then, that quantum phenomena, like all other physical phenomena we presently observe, refer to processes on some ``intermediate levels''. 
If the latter were the case, then ``the quantum'' would be an emergent phenomenon, i.e., based on some underlying sub-quantum physics, and limited by the boundary conditions of what one often terms ``macroscopic physics''.

In other words, like so many phenomena in the natural world, also the quantum phenomena may result from the interplay between various levels of description, rather than being reducible to one ``fundamental level'' only. In this sense, we shall here try to model the quantum as a well-coordinated \textit{emergent system}, where particle-like and wave-like phenomena are the result of both stochastic and regular dynamical processes. 
A prototype of such a system is well known from classical physics, viz., the ``bouncing droplets'' of Couder's group \cite{Couder.2005,Couder.2006single-particle,Protiere.2006,Eddi.2009,Fort.2010path-memory}, which in fact exhibit a whole series of phenomena reminiscent of quantum ones. 
Analogously, our group has in recent years attempted to model a quantum as a nonequilibrium steady-state maintained by a permanent throughput of energy. 
Specifically, we consider a ``particle'' as a ``bouncer'' whose oscillations are phase-locked with those of the energy-momentum reservoir of the surrounding ``vacuum'', i.e., the zero-point field. 
(The possible existence of a corresponding, underlying ``medium'' is \textit{a priori} independent of quantum theory. 
For similar views, compare, for example, the works of Hestenes~\cite{Hestenes.2009zitterbewegung}, Khrennikov~\cite{Khrennikov.2011prequantum}, Nelson~\cite{Nelson.1985quantum}, and Nottale~\cite{Nottale.2011scale}, as well as the approach of ``stochastic electrodynamics'' by de~la~Pe\~na and Cetto, and in particular a recent paper on the ``genesis of quantum nonlocality'' \cite{Pena.2011}.) 
In assuming that (part of) the zero-point fluctuations undergo regular oscillatory motion, where the latter is partly caused by and dynamically coupled to the oscillator's frequency $\omega$, we were able to derive fundamental elements of quantum theory from such a classical approach\footnote{A word of caution is appropriate here as to our use of the term ``classical'', which in the literature, in fact, can mean quite different things. Often, one finds that ``classical physics'' is understood as ``all physics with the exception of quantum theory''. When we use the expression ``classical'', we rather intend it to imply the ``updated'' present-day status of the classical paradigm of a century ago, i.e., including present-day statistical physics, nonequilibrium thermodynamics, and the like.
Vacuum fluctuations in our terminology thus refer to the statistical mechanics of a ``classical'' sub-quantum medium.
(For a similar use of terms, see Boyer \cite{Boyer.2010derivation}, for example.)}. 
In ref.~\cite{Groessing.2010explan} we showed how Planck's relation between the energy $E$ and the frequency $\omega$, $E=\hbar\omega$, can be derived from our sub-quantum physics, with Planck's (reduced) constant $\hbar$ indicating a universal angular momentum, and we also showed that with this relation alone one can derive the exact Schr\"odinger equation from classical physics \cite{Groessing.2008vacuum,Groessing.2009origin}. 
Moreover, also the stochastic element of the zero-point fluctuations enters decisively into our model, such that in effect we obtain an exact description of free quantum motion via a combination of the propagation of classical waves with diffusion due to stochastic sub-quantum mechanics. 
In particular, free quantum motion was shown to be exactly equivalent to anomalous (``ballistic'') diffusion, which is a) crucially dependent on the initial conditions, but b) time-reversible despite the usual irreversible nature of the more common forms of diffusion processes. \cite{Groessing.2010emergence}

To provide a suitable illustration, let us briefly discuss the following idealized scenario. Consider the straight path of a ``particle'' (e.g., an electron or a neutron) unaffected by any forces or apparatuses' boundary conditions. In other words, consider this path as a classical one. 
Then classical physics tells us that the Hamiltonian flow of such paths is formally accompanied by Huygens waves originating from the particle, such that their wave fronts are always orthogonal to the particle's path. 
If, for example, these waves then enter a slit system, Huygens' principle tells us how the diffraction at the slit is to be described. Now assume that said Huygens waves are not just mathematical tools, but physical waves excited by a bouncer's oscillatory motion in its surrounding medium. 
Then, in this highly idealized classical scenario, Huygens' principle would describe the actual physical process of waves centrally-symmetrically originating from the ``particle'' position. 
As the bouncing oscillation is phase-locked with that of the waves, one can imagine a steering mechanism in analogy to that observed by Couder's group \cite{Couder.2005,Couder.2006single-particle,Protiere.2006,Eddi.2009,Fort.2010path-memory} such that the bouncer is actually a ``walker'' along its classical path.

Now let us zoom in on such a system and consider the ``particle'' more closely, i.e., with statistical physics.
If the bouncer were the only one in the world, with no obstacles or other boundaries affecting its movement, then the latter would -- in its temporal mean -- always remain on a straight path, with regular Huygens waves emanating from it. 
But in the real world there are other objects and boundary conditions, which will affect the above-mentioned medium surrounding the ``particle'', i.e., contribute to a physical ``context'', just like the walker's paths of Couder's group are affected by objects such as a slit system. 
This means for our approach that the waves emanating from the bouncer and those impinging on it from the surrounding world will interact: some, more regular, parts may create standing waves, e.g., due to a synchrony between incoming and outgoing waves, but many will just act as ``disturbances'' of the regular path, i.e., factually act as stochastic influences due to the irregular part of the zero-point fluctuations. 
In other words, Huygens' principle will then strictly apply only to the combination of a ``walking'' motion of some velocity $\VEC{v}$ with centrally-symmetric diffusion waves' motion of velocity $\VEC{u}$ orthogonal to it. 
It will thus be only approximate, disturbed by that part of the accompanying diffusive process which is to be described by the corresponding velocity fluctuations $\delta\VEC{u}$. 
In fact, as we have shown for a Gaussian slit, the exact quantum mechanical result can be described as a combination of classical wave mechanics with the addition of a corresponding stochastic diffusion process \cite{Groessing.2010emergence}. 
In this paper, we intend to show that the same modelling procedure also perfectly applies to a double slit system, again providing the exact reproduction of the quantum mechanical results, albeit again relying on classical physics.

\section{Path excitation field and diffraction at a Gaussian slit}
\label{sec:path}

To begin with, we recall some of the basic results of our earlier work, including that on diffraction at a single Gaussian slit \cite{Groessing.2010emergence}. We claim for the system of a bouncer of frequency $\omega$ embedded in a stochastic (``zero-point'') environment that its \textit{average total energy} is given by 
the average ``total'' energy $\meanx{\hbar\omega}$ of the bouncer itself plus a kinetic energy term due to momentum changes $\VEC{p}_u =: m\VEC{u}$ which it receives from or gives off to the environment:
\begin{equation} \label{eq:2.1} 
  \meanx{E_{\rm tot}} = \meanx{\hbar \omega}+\frac{\meanx{p_u^2}}{2m} = \text{const}, 
\end{equation} 
where the averaging (as denoted by the bars) is defined in $n$-dimensional configuration space as
\begin{equation} \label{eq:2.3} 
  \meanx{p_u^2}:=\int P p_u^2 \d^{n} x . 
\end{equation} 
$P=P(\VEC{x},t)$ refers to the probability density of some relevant distribution. For our model system, the latter will be introduced below as a solution of a generalized (``anomalous'') diffusion equation,
\begin{equation} \label{eq:2.4} 
  \frac{\partial}{\partial t} P(x,t) = D(t)\frac{\partial^2}{\partial x^2} P(x,t), 
\end{equation} 
with $D(t)$ being a time-dependent diffusion coefficient. 
As can be seen from \eqref{eq:2.1}, the momentum changes can be either positive or negative, and actually will \textit{on the average} balance each other, since they are \textit{a priori} unbiased. 
The deeper reason for this balancing in our model is due to the fact that we consider the bouncer/walker to be a steady-state system in the sense that its ``total'' energy $\hbar\omega$ is maintained over times $t\gg1/\omega$ by the permanent throughput of kinetic energy $p_u^2/2m$. 
In other words, to maintain the steady-state, during the intervals of the average order of $t\simeq 1/\omega$, there will both be an absorption of a momentum $p_u = mu$ and a release of the same amount, $p_u = -mu$.

In our earlier papers \cite{Groessing.2008vacuum,Groessing.2009origin,Groessing.2010emergence}, or alternatively in \cite{Garbaczewski.1992derivation}, for example, one derives from classical physics the expression for $\VEC{u}$ as
\begin{equation} \label{eq:2.5a}
	\VEC{u} = -D \frac{\nabla P}{P},
\end{equation}
with the diffusion constant $D$. Noting that in reference~\cite{Groessing.2010explan}, Planck's (reduced) constant $\hbar$ is given  by a basic angular momentum,
\begin{equation} \label{eq:2.2} 
  \hbar = m\omega r^2, 
\end{equation} 
and if $r$ in \eqref{eq:2.2} is identified with the usual expression for a diffusion length, $r=\sqrt{2D/\omega}$, then one obtains the well-known diffusivity
\begin{equation} \label{eq:2.5b} 
  D = \frac{\hbar}{2m}.
\end{equation}

So, apart from the ordinary particle current $\VEC{J}(\VEC{x},t) = P(\VEC{x},t)\VEC{v}$, we are dealing with two additional, yet opposing, currents $\VEC{J}_u$, which are on average 
orthogonal to $\VEC{J}$ \cite{Groessing.2010explan,Groessing.2008vacuum,Groessing.2009origin,Groessing.2010emergence}, and which are the emergent outcome from the presence of numerous corresponding velocities
\begin{equation} \label{eq:2.7} 
  \VEC{u}_\pmbracket = \mpbracket\frac{\hbar}{2m}\frac{\nabla P}{P} .
\end{equation}

We denote with $\VEC{u}_+$ and $\VEC{u}_-$, respectively, the two opposing tendencies of the diffusion process.
In the reference frame of a single free particle, and starting at $t=0$ at the center of the distribution $P$,
the averages obey
\begin{equation} \label{eq:2.7a} 
  \meanx{\VEC{u}}_-(\VEC{x},t) = -\meanx{\VEC{u}}_+(-\VEC{x},t) .
\end{equation}

Now let us consider an experimental setup with a particle source. To describe the velocity distribution, we introduce a velocity field with average velocity $\meanx{\VEC{v}}$, and amplitudes $R(\VEC{x},t)$.
We refer to their intensities $P=R^2$ as the solutions of a diffusion equation such as \eqref{eq:2.4}. These typically appear in the form of a Gaussian distribution $P(\VEC{x},t)$ of possible particle locations $\VEC{x}$, even if there is only one particle at a time emerging from the corresponding ``Gaussian slit'',
in one dimension for simplicity,
\begin{equation} \label{eq:2.8} 
  P(x,t) = \frac{1}{\sqrt{2\pi} \, \sigma} \e^{-\frac{(x-x_{0})^2}{2\sigma^2}} , 
\end{equation} 
with the usual variance $\sigma^2 = \meanx{\left(\Delta x\right)^2} = \meanx{\left(x-x_{0} \right)^2}$, and where we choose $x_0=0$ further on. 
Regarding $\VEC{u}$, even in this scenario of one-particle-at-a-time, we deal with an \textit{ensemble} of velocity vectors $\VEC{u}_\alpha(t)$ representing hypothetical motions on the sub-quantum level in a small volume around $\VEC{x}$, whose mean value will be given by
\begin{equation} \label{eq:2.9} 
  \VEC{u}(\VEC{x},t)=\frac{1}{N(\VEC{x}, t)}\sum_{\alpha=1}^{N(\VEC{x}, t)} \VEC{u}_\alpha(t) . 
\end{equation} 
Here, the (typically very large) number $N$ refers to the number of possible path directions of the bouncer due to the existence of the wave-like excitations of the zero-point field.
This may be reminiscent of Feynman's picture of photons \textit{virtually} probing every possible path in an experimental setup, but in our case it is a configuration of \textit{real} wave-like excitations, which in a resulting Brownian-type motion guide the bouncer along its path of average velocity $\VEC{u}$. 
Again, we note that here we discuss not only a passive guidance of the ``particle'' by the surrounding wave configurations, but point out also the very active role of the ``particle'' in (partly) \textit{creating} said wave configurations due to the effects of its bouncing. 
To account for \eqref{eq:2.7}, we split up $\VEC{u}(\VEC{x},t)$ according to
\begin{equation}  \label{eq:3.6}
  \VEC{u}(\VEC{x},t) = \frac{1}{2N}\left[\sum_{\alpha=1}^N \VEC{u}_{\alpha,+} + \sum_{\alpha=1}^{N} \VEC{u}_{\alpha,-} \right] 
      = \frac{1}{2} \left[\VEC{u}_+ + \VEC{u}_-\right],
\end{equation}
thus reflecting the isotropy of the diffusion process.
Still, the uncontrollable and possibly unknowable velocity field $\VEC{u}$ representing the Brownian motion of the bouncer may not be operational, but when we take the average according to the rule \eqref{eq:2.3}, we obtain a ``smoothed-out'' \textit{average velocity field} 
\begin{equation}  \label{eq:2.10}
  \meanx{\VEC{u}(\VEC{x},t)} = \int P\VEC{u}(\VEC{x},t) \d^nx ,
\end{equation}
which is all that we need for our further considerations.
Similarly, based on the fact that we have an initial Gaussian distribution of velocity vectors $\VEC{v}(\VEC{x},t)$, we define an average velocity field $\meanx{\VEC{v}}$ of the wave propagation as
\begin{equation}  \label{eq:2.11}
  \meanx{\VEC{v}(\VEC{x},t)} = \int P\VEC{v}(\VEC{x},t) \d^nx ,
\end{equation}
and we here just note what we have discussed extensively in previous papers \cite{Groessing.2008vacuum,Groessing.2009origin,Groessing.2010emergence,Groessing.2011dice}, i.e., that there is an average orthogonality between the two velocity fields,
$\VEC{u}$ and $\VEC{v}$,
\begin{equation}  \label{eq:2.12}
  \meanx{\VEC{v}\cdot\VEC{u}} = \int P\VEC{v}\cdot\VEC{u} \d^nx = 0 .
\end{equation}
This holds for a variety of reasons, one of them simply being that $\VEC{u}$ and $\VEC{v}$ are linearly uncorrelated.

In effect, then, the combined presence of both velocity fields $\VEC{u}$ and $\VEC{v}$ can be denoted as a \textit{path excitation field}: via diffusion, the bouncer in its interaction with already existing wave-like excitations of the environment creates a ``heated-up'' thermal ``landscape'', which can also be pictured by interacting wave configurations all along between source and detector of an experimental setup. 
Recall that our prototype of a ``walking bouncer'', i.e., from the experiments of Couder's group, is always driven by its interactions with a superposition of waves emitted at the points it visited in the past. Couder et~al. denote this superposition of in-phase waves the ``path memory'' of the bouncer \cite{Fort.2010path-memory}.
This implies, however, that the bouncers at the points visited in ``the present'' necessarily create new wave configurations which will form the basis of a ``path memory'' in the future.
In other words, the wave configurations of the past determine the bouncer's path in the present, whereas its bounces in the present co-determine the wave configurations at any of the possible locations it will visit in the future.
Therefore, we call the latter configurations the \textit{path excitation field}, which may also be described as ``heated-up'' thermal field.
As in the coupling of an oscillator with classical diffusion, diffusion wave fields arise with instantaneous field propagation \cite{Groessing.2009origin,Mandelis.2001structure}, one has elements of the whole setup which may be nonlocally oscillating (``breathing'') in phase. 
This means that the Gaussian solving \eqref{eq:2.4} does represent a nonlocal path excitation field in that it is a physically existing and effective entity responsible for where the bouncing ``particle'' can possibly go. 
As we have shown \cite{Groessing.2010emergence}, one can in this classical framework, along with the time-dependence of $D(t)$, effectively and easily describe the (sub-)quantum physics of diffraction at a Gaussian slit, which we now briefly recapitulate.

At first we note that Eq.~\eqref{eq:2.1} is an \textit{average} energy conservation law only. This means that apart from the momentum changes $\VEC{p}_u = \pm m\VEC{u}$ discussed so far, also variations in $\VEC{p}_u$ will have to be taken into account, and thus also variations in the ``particle energy'' $\hbar\omega$. 
If for the latter one just considers its kinetic energy term, $mv^2/2$, then said variations will lead to exchanges of velocity/momentum terms providing the net balance
\begin{equation}  \label{eq:2.13}
  m\delta\VEC{v} = m\delta\VEC{u} .
\end{equation}
Thus, the \textit{average} kinetic energy variation will be determined, once an initial condition is defined:
\begin{equation}  \label{eq:2.14}
  \meanx{\delta E_{\rm kin}(t)} = \meanx{\delta E_{\rm kin}(0)} = \text{const}.
\end{equation}
In other words, one starts with the initial average squared momentum change, which is determined by the characteristics of the particle source (to be shown below),
\begin{equation}  \label{eq:2.15}
  \left.\meanx{p_u^2} = m^2\,\meanx{u^2}\right|_{t=0} =: m^2u_0^2 ,
\end{equation}
relative to which all further momentum balances have to be carried out. Then, \eqref{eq:2.14} reads
\begin{equation}  \label{eq:2.16}
  \meanx{\delta E_{\rm kin}(0)} = \frac{m}{2}u_0^2 = \meanx{\delta E_{\rm kin}(t)} = \frac{m}{2}\left(\meanx{u^2} + \meanx{(\delta u)^2}\right) .
\end{equation}
Using the expression \eqref{eq:2.7} for $\VEC{u}$, one obtains with Eqs.~\eqref{eq:2.8} and \eqref{eq:2.5b}, and with $\meanx{(\nabla\ln P)^2} = -\meanx{\nabla^2\ln P}$, that
\begin{equation}  \label{eq:2.17}
  u_0^2 = \frac{D^2}{\sigma_0^2} = \meanx{u^2} + \meanx{(\delta u)^2} = \frac{D^2}{\sigma^2} +\meanx{(\delta u)^2} ,
\end{equation}
where as usual $\sigma=\sigma(t)=\sqrt{\,\meanx{x^2}}$ for $x_0(t=0)=0$, and $\sigma_0 = \sigma(t=0)$.
One can view the Gaussian distribution $P$ of kinetic energy also as a sort of ``heat accumulation'', which has its maximum at the center. Considering now the application of momentum fluctuations (up to second order) to a particle with initial ($t=0$) distance $x(0)$ from said center, with the fluctuation term for $t>0$ defined as 
$\VEC{p}_u\pm\delta\VEC{p}_u=\pm m(\VEC{u}\pm\delta\VEC{u})$, one will find said particle at time $t$ at the location
\begin{equation}  \label{eq:2.18}
  \VEC{x}(t) = \VEC{x}(0) \pm \left(\VEC{u} \pm \delta\VEC{u}\right)t .
\end{equation}
Squaring Eq.~\eqref{eq:2.18} and forming the r.m.s.\ then provides with Eq.~\eqref{eq:2.16} and average orthogonality \eqref{eq:2.12}, and thus also $\meanx{\VEC{u}\cdot\VEC{x}(0)}=0$ \cite{Groessing.2010emergence},
\begin{equation}  \label{eq:2.19}
  \meanx{x^2}(t) = \meanx{x^2}(0) + \left[\meanx{u^2} + \meanx{(\delta u)^2}\right] t^2 = \meanx{x^2}(0) + u_{0}^2 t^2 . 
\end{equation}
(Note that we could have included still higher order terms of the fluctuations, but due to the conservation condition \eqref{eq:2.14}, Eq.~\eqref{eq:2.19} results also for any number of higher orders.) Comparing with Eq.~\eqref{eq:2.17} also provides the time evolution of the wave packet's variance as
\begin{equation}  \label{eq:2.20}
  \sigma^2 = \sigma_{0}^2 \left(1+\frac{D^2 t^2}{\sigma_{0}^{4}} \right).
\end{equation}
As the spreading ratio $\sigma/\sigma_0$ for the wave packet is independent of $x$, and thus not only valid for the particular point $x(t)=\sigma(t)$, one can generalize the square root of Eq.~\eqref{eq:2.20} for all $x$, thus providing
\begin{equation}  \label{eq:2.21}
  x(t) = x(0)\frac{\sigma}{\sigma_{0}}, \quad \text{where } \; \frac{\sigma}{\sigma_{0}} 
      = \sqrt{1+\frac{D^2 t^2}{\sigma_{0}^{4}}} \;. 
\end{equation}
Thus, one obtains for ``smoothed-out'' trajectories (i.e., averaged over a very large number of Brownian motions) a sum over a deterministic and a fluctuations term, respectively, i.e., with 
Eq.~\eqref{eq:2.5b},
\begin{equation}  \label{eq:2.22}
  x_{\rm tot} (t)= v t + x(t) = v t + x(0)\frac{\sigma}{\sigma_{0}} 
      = v t + x(0)\sqrt{1+\frac{\hbar^2 t^2}{4m^2 \sigma_{0}^{4}}} .
\end{equation}
Moreover, one thus derives with Eq.~\eqref{eq:2.17} the average velocity field of a Gaussian wave packet as
\begin{equation}  \label{eq:2.23}
  v_{\rm tot}(t) = v(t) + \frac{\d x(t)}{\rm \d t} = v(t)+ \left[x_{\rm tot}(t) - v t \right]\frac{u_0^2 t}{\sigma^2} .
\end{equation}

Note that Eqs.~\eqref{eq:2.20}-\eqref{eq:2.23} are derived solely from statistical physics. Still, they are in full accordance with quantum theory, and in particular with Bohmian trajectories \cite{Holland.1993}. Note also that one can rewrite Eq.~\eqref{eq:2.19} such that it appears like a linear-in-time formula for Brownian motion,
\begin{equation}  \label{eq:2.24}
  \meanx{x^2} = \meanx{x^2(0)} + D(t)\,t ,
\end{equation}
where a time dependent diffusivity     
\begin{equation}  \label{eq:2.25}
  D(t) = u_0^2 \, t = \frac{\hbar^2}{4m^2\sigma_0^2}\,t
\end{equation}
characterizes Eq.~\eqref{eq:2.24} as \textit{ballistic diffusion}.
The appearance of a time-dependent $D(t)$ is easy to understand. 
The diffusivity is changed over time, because the ``particle's'' thermal environment changes: With the ``heat'' initially concentrated within the narrow spatial constraints determined by $\sigma_0$ of the source (``Gaussian slit''), $D(t)$ must become larger with time because of the gradually lower heat concentration due to dissipation into the unconstrained environment.
(Note that a similar scenario was suggested by Garbaczewski~\cite{Garbaczewski.1992derivation}; others are presently intensively discussed in the context of the so-called ``superstatistics'' \cite{Beck.2008}.
Moreover, ballistic diffusion has also been measured recently in experiments realizing quantum walks, see, e.g., \cite{Regensburger.2011zitterbewegung} and references therein.)
This makes it possible to simulate the dispersion of a Gaussian wave packet on a computer by simply employing coupled map lattices for classical diffusion, with the diffusivity given by Eq.~\eqref{eq:2.25}. (For detailed discussions, see refs.~\cite{Groessing.2010emergence} and \cite{Groessing.2011dice}.)

\section{An explanation of interference effects in the double slit experiment}
\label{sec:quantum}

With the essentials of Gaussian dispersion at our disposal, it is straightforward to now also describe and explain quantum interference with our approach. We choose a textbook scenario in the form of the calculation of the intensity distribution and the particle trajectories in an electron interferometer. As we are also interested in the trajectories, we refer to, and compare our results with, the well-known work by Philippidis et~al. \cite{Philippidis.1979quantum}, albeit in the form as presented by Holland \cite{Holland.1993}.

We choose similar initial situations as in \cite{Holland.1993}, i.e., electrons (represented by plane waves in the forward $y$-direction) from a source passing through ``soft-edged'' slits $1$ and $2$ in a barrier (located along the $x$-axis) and recorded at a screen. In our model, we therefore note two Gaussians representing the totality of the effectively ``heated-up'' path excitation field, one for slit $1$ and one for slit $2$, whose centers have the distances $+X$ and $-X$ from the plane spanned by the source and the center of the barrier along the $y$-axis, respectively. 
We recall from classical wave mechanics that the plane-wave superposition of two amplitudes $R_i=\sqrt{P_i}$, $i=1$ or $2$, with resulting amplitude $R$ and wave number $\VEC{k}$ provides
\begin{equation}  \label{eq:3.0}
  R\VEC{k} = R_1\VEC{k}_1 + R_2\VEC{k}_2  ,
\end{equation}
such that for $|\VEC{k}_1|=|\VEC{k}_2|=|\VEC{k}|$, one obtains (with hats further on denoting unit vectors) the total intensity
\begin{equation}  \label{eq:3.0a}
  P_{\rm tot} := R^2 = \left| R_1\VEC{\hat{k}}_1 + R_2\VEC{\hat{k}}_2 \right|^2 
      = R_1^2 + R_2^2 + 2R_1R_2\cos\varphi 
      = P_1+P_2+2\sqrt{P_1P_2}\cos\varphi .
\end{equation}
The $x$-components of the centroids' motions from the two alternative slits $1$ and $2$, respectively, are given by the ``particle'' velocity components 
\begin{equation}  \label{eq:3.1}
  v_x = \pm \frac{\hbar}{m} \, k_x ,
\end{equation}
respectively, such that the relative group velocity of the Gaussians spreading into each other is given by $\Delta v_x=2v_x$. 
However, in order to calculate the phase difference $\varphi$ descriptive of the interference term of the intensity distribution \eqref{eq:3.0a}, one must take into account also the wave packet dispersion as described in the previous Chapter. 
Thus, one obtains with Eq.~\eqref{eq:2.23} the total relative velocity of the two Gaussians as 
\begin{equation}  \label{eq:3.1a}
  \Delta v_{{\rm tot}, x} = 2\left[ v_x - (X + v_x t) \frac{\hbar^2}{4m^2} \frac{t}{\sigma^2\sigma_0^2} \right] .
\end{equation}
Therefore, the total phase difference between the two possible paths (i.e., through either slit) becomes
\begin{equation}  \label{eq:3.2}
  \varphi = \frac{1}{\hbar}(m\Delta v_{{\rm tot}, x} \, x) = 2mv_x \frac{x}{\hbar} - (X + v_x t) x \frac{\hbar}{2m} \frac{t}{\sigma^2\sigma_0^2} ,
\end{equation}
where the last term also reads as
\begin{equation}  \label{eq:3.3}
  - (X + v_x t) x \frac{1}{D} \frac{u_0^2 t}{\sigma^2} = - (X + v_x t) x \frac{1}{D} \frac{\dot{\sigma}}{\sigma} .  
\end{equation}
The Gaussians $P_1$ and $P_2$ for the corresponding slits are given as
\begin{equation}  \label{eq:3.3b}
  P_1(x,t) = \frac{1}{\sqrt{2\pi\sigma^2}} \e^{-[x - (X + v_x t)]^2/2\sigma^2} ,
\end{equation}
and
\begin{equation}  \label{eq:3.3c}
  P_2(x,t) = \frac{1}{\sqrt{2\pi\sigma^2}} \e^{-[x + (X + v_x t)]^2/2\sigma^2} .
\end{equation}
With equal amplitudes $R=R_i=\sqrt{P_i}$, for $i=1,2$, of the Gaussians, and with normalization constant $N$, we thus obtain the usual interference pattern in the form of the intensity distribution:
\begin{equation}  \label{eq:3.4}
    P_{\rm tot}(x,t) = R^2N^2 \frac{1}{\sqrt{2\pi\sigma^2}}\e^{-[x^2 + (X + v_xt)^2]/2\sigma^2} \left\{e^{x(X + v_xt)/\sigma^2} + e^{-x(X + v_xt)/\sigma^2} + 2\cos\varphi \right\} ,
\end{equation}
with $\varphi$ given by Eq.~\eqref{eq:3.2}. The term inside the curly brackets of Eq.~\eqref{eq:3.4} describes the interference fringes whose ``dark'' nodes are at the locations well-known from textbooks,
\begin{equation}  \label{eq:3.5}
  x = (n + \frac{1}{2})\pi / k_x ,
\end{equation}
if, according to Eq.~\eqref{eq:3.3}, either $\dot{\sigma}/\sigma$ is negligible (i.e., $\sigma(t)\simeq \text{const.}$), or under the specific time-conditions $X=-v_xt$, as the wave packets must approach each other ($v_x<0$) for $t>0$. 
For a discussion of further aspects including, for example, those on the Fraunhofer limit, we refer the reader to the details provided in Holland's book \cite{Holland.1993}, which equally result from our Eq.~\eqref{eq:3.4}.

\begin{figure}[!t]
 \center
		\includegraphics[width=130mm,height=130mm]{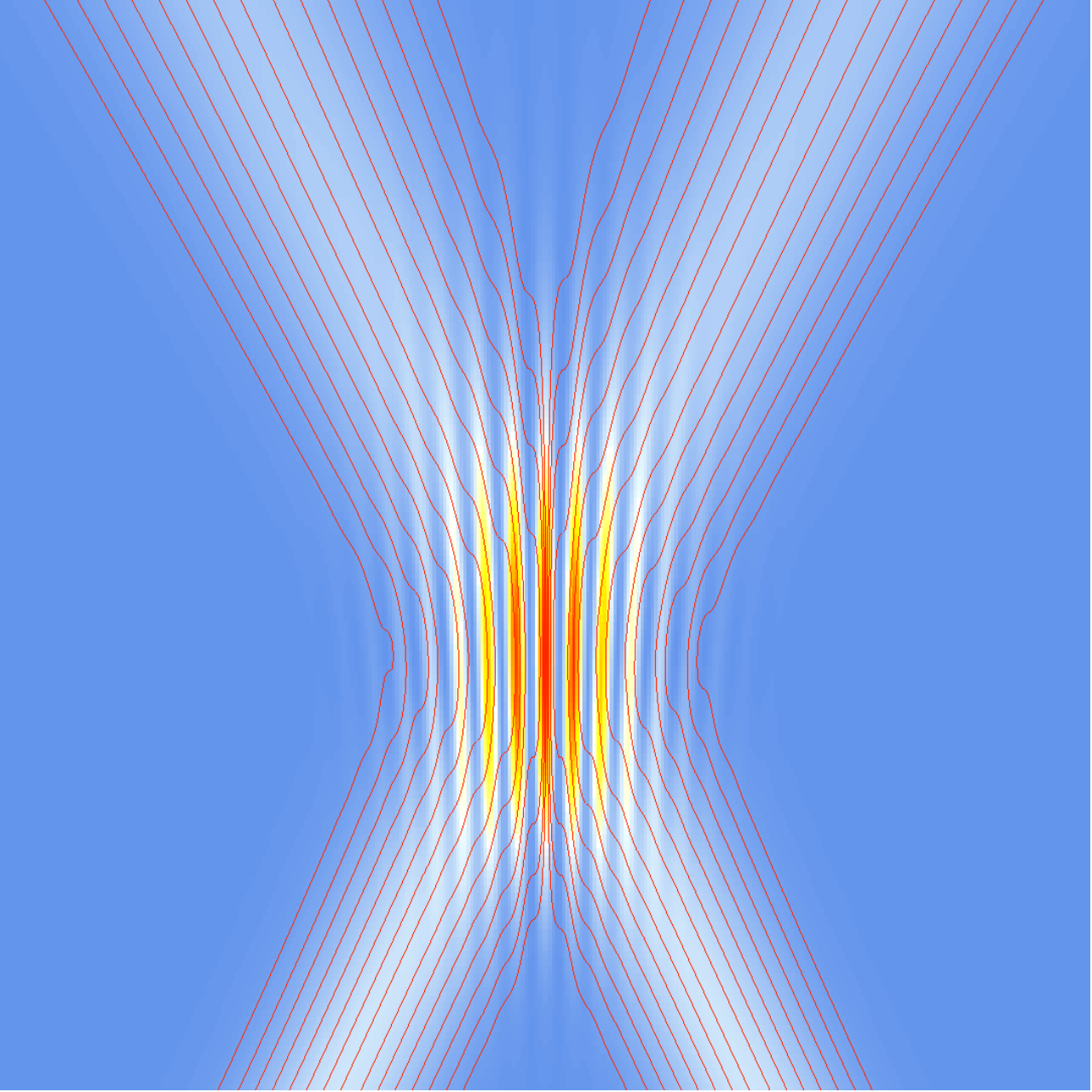}
		\caption{Classical computer simulation of the interference pattern: intensity distribution with increasing intensity from white through yellow and orange, with trajectories (red) for two Gaussian slits, and with small dispersion (evolution from bottom to top; $v_{x,1}=-v_{x,2}$).
		The trajectories follow a ``no crossing'' rule: particles from the left slit stay on the left side and \textit{vice versa} for the right slit. 
		This feature is explained here by a sub-quantum build-up of kinetic (heat) energy acting as an emergent repellor along the symmetry line. }
		\label{fig:2}
\end{figure}

Fig.~\ref{fig:2} depicts the interference of two beams emerging from Gaussian slits established with the aid of a simulation as in ref.~\cite{Groessing.2010emergence}, where we simulate diffusion with a time-dependent diffusivity $D(t)$.
To account for interference, we simply follow the classical rule for the intensities \eqref{eq:3.0a}, with $\varphi$ from Eq.~\eqref{eq:3.2}.
The trajectories are the flux lines obtained by choosing a set of equidistant initial points at $y=0$.
Two adjacent flux lines thereby define regions of constant difference $\Delta P$.

Just as with the dispersion of a single Gaussian, we want to stress that also the quantum interference pattern \eqref{eq:3.4} has been derived here without the use of quantum mechanics, but solely on the basis of classical physics as it was outlined in the footnote of the introduction. 
Exploiting our concept of the ``path excitation field'', we have for this derivation implicitly used the assumption of velocity fields 
$\VEC{u}(\VEC{x},t)$ and $\VEC{v}(\VEC{x},t)$, respectively, which have entered the expression of a Gaussian's average velocity field, Eq.~\eqref{eq:2.23}, and which in turn were shown to essentially contribute to the interference pattern \eqref{eq:3.4}. 
In what follows, we now want to make the use of these velocity fields more explicit, i.e., we shall now concentrate on understanding the emerging particle trajectories during quantum interference on the basis of our classical velocity fields.

\section{Geometric meaning of the path excitation field and its translation into quantum language}
\label{sec:geometric}

\begin{figure}[!t]
	\center
	\includegraphics[width=120mm,height=105mm]{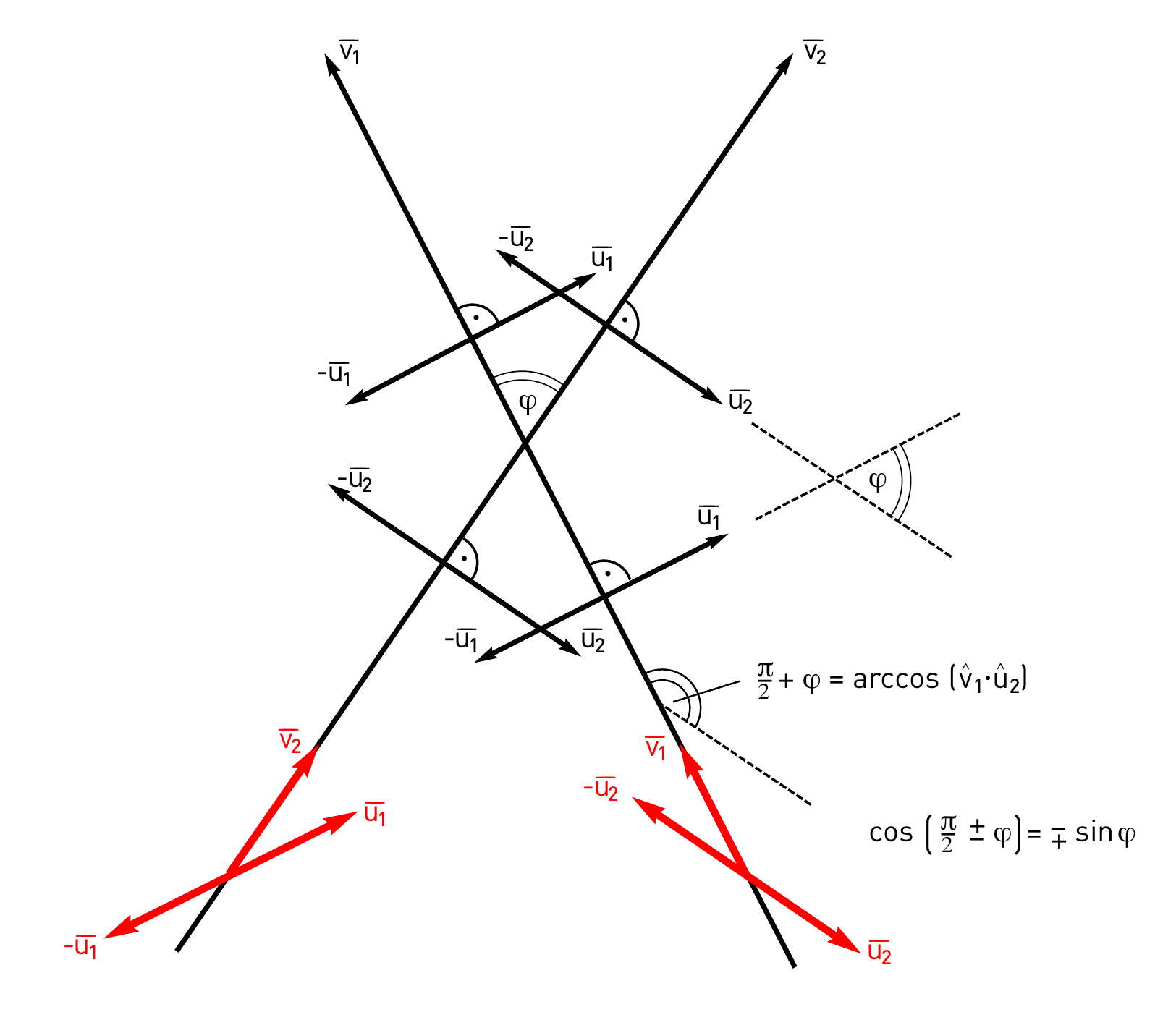}  
	\caption{Two-dimensional schematic of the path excitation field behind a double slit.
	$\meanx{u}_i$ and $\meanx{v}_i$, $i=1$ or $2$, denote the average directions of the velocity fields involved, with $\meanx{v}_i$ referring to single free ``particle'' velocities (i.e., either through slit 1 or slit 2), and the $\meanx{u}_i$ referring to additional diffusion velocities.
	The red arrows indicate how the mixing of velocity fields $\meanx{v}_i$ and $\meanx{u}_j$, with $i\neq j$, produces the terms proportional to $\sin\varphi$ in Eq.~\eqref{eq:3.12}.
	Upon superposition of the two channels, according to Eq.~\eqref{eq:3.12}, one obtains the final particle trajectories whose emergent behavior is characterized by a ``no crossing'' rule (see Fig.~\ref{fig:2}). }
	\label{fig:1}
\end{figure}

Let us consider a single, classical ``particle'' (``bouncer'') following the propagation of a set of waves of equal amplitude $R$, each representing one of $i$ possible alternatives according to our principle of path excitation.
We now concentrate on the specific role of the velocity fields $\VEC{u}$, which were present in Chapter~\ref{sec:quantum} only implicitly in the expressions \eqref{eq:3.0a} and \eqref{eq:3.2}.
To describe the required details, each path $i$ be occupied by a Gaussian wave packet with a ``forward'' momentum $\VEC{p}_i=\hbar\VEC{k}_i=m\VEC{v}_i$.
Moreover, due to the stochastic process of path excitation, the latter has to be represented also by a large number $N$ of consecutive Brownian shifts, $\VEC{p}_{u,\alpha}=m\VEC{u}_\alpha$.
Recalling \eqref{eq:3.6}, one has
\begin{equation}  \label{eq:4.1}
  \VEC{u}(\VEC{x},t) = \frac{1}{2N}\left[\sum_{\alpha=1}^N \VEC{u}_{\alpha,+} + \sum_{\alpha=1}^{N} \VEC{u}_{\alpha,-} \right] 
      = \frac{1}{2} \left[\VEC{u}_+ + \VEC{u}_-\right].
\end{equation}
This provides for the case of interference at a double slit the total velocity field (with indices $i=1$ or $2$ referring to the two slits)
\begin{equation}  \label{eq:3.8}
	\meanx{\VEC{v}}_{\rm tot} = \meanx{\VEC{v}}_{{\rm tot},1} + \meanx{\VEC{v}}_{{\rm tot},2}
		:= \meanx{\VEC{v}}_1 
		+ \frac{\meanx{\VEC{u}}_{1+}}{2} 
		+	\frac{\meanx{\VEC{u}}_{1-}}{2}
		+ \meanx{\VEC{v}}_2
		+ \frac{\meanx{\VEC{u}}_{2+}}{2}
		+ \frac{\meanx{\VEC{u}}_{2-}}{2}.
\end{equation} 
With two Gaussian distributions $P_1=R_1^2$ and $P_2=R_2^2$ as given above, one has
\begin{equation}  \label{eq:3.9}
	R_{\rm tot}\meanx{\VEC{v}}_{\rm tot}
		= R_1\meanx{\VEC{v}}_{{\rm tot},1}
		+ R_2\meanx{\VEC{v}}_{{\rm tot},2} \;,
\end{equation}
which approximates the addition of amplitudes at the crossing point of plane waves.
To obtain an expression for $P_{\rm tot}=R_{\rm tot}^2$ and for the relevant velocity components in the interference region, we derive from Eq.~\eqref{eq:3.9}
\begin{equation}  \label{eq:3.10}
	R_{\rm tot} = 
		\left[R_1\left(\meanx{\VEC{v}}_1 
		+ \frac{\meanx{\VEC{u}}_{1+}}{2} 
		+	\frac{\meanx{\VEC{u}}_{1-}}{2}\right)
		+ R_2\left(\meanx{\VEC{v}}_2
		+ \frac{\meanx{\VEC{u}}_{2+}}{2}
		+ \frac{\meanx{\VEC{u}}_{2-}}{2}\right)\right]
		\frac{\hat{\VEC{v}}_{\rm tot}}{|\meanx{\VEC{v}}_{\rm tot}|}.
\end{equation}
Taking into account the conservation of the ``particle momentum'' in both channels, we have 
$|\meanx{\VEC{v}}_{\rm tot}| = |\meanx{\VEC{v}}_{{\rm tot},1}| = |\meanx{\VEC{v}}_{{\rm tot},2}|$, leading to
\begin{equation}  \label{eq:3.10a}
	R_{\rm tot} = 
		\left[R_1\left(\hat{\VEC{v}}_1 
		+ \frac{\hat{\VEC{u}}_{1+}}{2} 
		+	\frac{\hat{\VEC{u}}_{1-}}{2}\right)
		+ R_2\left(\hat{\VEC{v}}_2
		+ \frac{\hat{\VEC{u}}_{2+}}{2}
		+ \frac{\hat{\VEC{u}}_{2-}}{2}\right)\right]
		\hat{\VEC{v}}_{\rm tot}.
\end{equation}
To help with the bookkeeping, Fig.~\ref{fig:1} displays all the relevant vectors and some of their included angles.
With Eq.~\eqref{eq:2.7a}, and setting $\VEC{u}_{1+} \to \VEC{u}_{1}$ and $\VEC{u}_{1-} \to -\VEC{u}_{1}$, one obtains the total average current
\begin{equation}  \label{eq:3.11}
  \begin{array}{r@{}r@{\;}c@{\;}r@{\;}c@{\;}r@{\;}l@{}}
		 \multicolumn{7}{l}{\meanx{\VEC{J}}_{\rm tot} = 
					R_{\rm tot}^2\meanx{\VEC{v}}_{\rm tot} = 
					R_1^2 \meanx{\VEC{v}}_1 + R_2^2 \meanx{\VEC{v}}_2 +}  \\ [1.5ex]
		 + R_1R_2 \biggl\{ & \displaystyle
            \left(\meanx{\VEC{v}}_1 + \meanx{\VEC{v}}_2\right)\cos(\hat{\VEC{v}}_1,  \hat{\VEC{v}}_2) 
          &+& \displaystyle \left(\meanx{\VEC{v}}_1 + \frac{\meanx{\VEC{u}}_2}{2}\right)\cos(\hat{\VEC{v}}_1,  \hat{\VEC{u}}_2) 
          &-& \displaystyle \left(\meanx{\VEC{v}}_1 - \frac{\meanx{\VEC{u}}_2}{2}\right)\cos(\hat{\VEC{v}}_1,  \hat{\VEC{u}}_2) & \biggr. \\ [2ex]
     +& \displaystyle
				    \left(\frac{\meanx{\VEC{u}}_1}{2} + \meanx{\VEC{v}}_2\right)\cos(\hat{\VEC{u}}_1, \hat{\VEC{v}}_2)
					&-& \displaystyle \left(-\frac{\meanx{\VEC{u}}_1}{2} + \meanx{\VEC{v}}_2\right)\cos(\hat{\VEC{u}}_1, \hat{\VEC{v}}_2)
          &+& \displaystyle \left(\frac{\meanx{\VEC{u}}_1}{2} + \frac{\meanx{\VEC{u}}_2}{2}\right)\cos(\hat{\VEC{u}}_1, \hat{\VEC{u}}_2) & \biggl.\biggr.\\ [2ex]
     -& \displaystyle
				    \left(\frac{\meanx{\VEC{u}}_1}{2} - \frac{\meanx{\VEC{u}}_2}{2}\right)\cos(\hat{\VEC{u}}_1, \hat{\VEC{u}}_2)
					&-& \displaystyle \left(-\frac{\meanx{\VEC{u}}_1}{2} + \frac{\meanx{\VEC{u}}_2}{2}\right)\cos(\hat{\VEC{u}}_1, \hat{\VEC{u}}_2)
          &+& \displaystyle \left(-\frac{\meanx{\VEC{u}}_1}{2} - \frac{\meanx{\VEC{u}}_2}{2}\right)\cos(\hat{\VEC{u}}_1, \hat{\VEC{u}}_2) & \biggl.\biggr\} ,
  \end{array} 
\end{equation}
and thus finally
\begin{equation}  \label{eq:3.12}
  \meanx{\VEC{J}}_{\rm tot} = P_1\meanx{\VEC{v}}_1 + P_2\meanx{\VEC{v}}_2 
      + \sqrt{P_1P_2}\left(\meanx{\VEC{v}}_1 + \meanx{\VEC{v}}_2\right)\cos\varphi
      + \sqrt{P_1P_2}\left(\meanx{\VEC{u}}_1 - \meanx{\VEC{u}}_2\right)\sin\varphi .
\end{equation}
Note that Eq.~\eqref{eq:3.12}, upon the identification of $\meanx{\VEC{u}}_i=-\frac{\hbar}{m}\frac{\nabla R_i}{R_i}$ from Eq.~\eqref{eq:2.7} and 
with $P_i = R_i^2$, turns out to be in perfect agreement with a comparable ``Bohmian'' derivation~\cite{Holland.1993,Sanz.2008trajectory}.
The formula for the particle trajectories, then, simply results from Eq.~\eqref{eq:3.11}, i.e.,
\begin{equation}  \label{eq:3.13}
  \meanx{\VEC{v}}_{\rm tot} = \frac{1}{P_{\rm tot}}\meanx{\VEC{J}}_{\rm tot} \;,
\end{equation}
with $P_{\rm tot}$ given by Eq.\eqref{eq:3.0a} and $\meanx{\VEC{J}}_{\rm tot}$ given by Eq.~\eqref{eq:3.12}.

In Fig.~\ref{fig:2} one can observe a basic characteristic of the (averaged) particle trajectories, which, just because of the averaging, are identical with the Bohmian trajectories. 
To fully appreciate this surprising characteristic, we remind the reader of the severe criticism of Bohmian trajectories as put forward by Scully and others 
(see \cite{Scully.1998bohm}, and references therein.) 
The critics claimed that Bohmian trajectories would have to be described as ``surreal'' ones because of their apparent violation of momentum conservation. 
In fact, due to the ``no crossing'' rule for Bohmian trajectories in Young's double slit experiment, for example, the particles coming from, say, the right slit (and expected at the left part of the screen if momentum conservation should hold on the corresponding macro-level) actually arrive at the right part of the screen (and \textit{vice versa} for the other slit). 
In Bohmian theory, this ``no crossing'' rule is due to the action of the non-classical quantum potential, such that, once the existence of a quantum potential is accepted, no contradiction arises and the trajectories may be considered ``real'' instead of ``surreal''.

Here we can note that in our sub-quantum approach an explanation of the ``no crossing'' rule is even more straightforward and actually a consequence of a detailed \textit{microscopic momentum conservation}. 
As can be seen in Fig.~\ref{fig:2}, the (Bohmian) trajectories are repelled from the central symmetry line. 
However, in our case this is only implicitly due to a ``quantum potential'', but actually due to the identification of the latter with a kinetic (rather than a potential) energy: 
As has already been stressed in \cite{Groessing.2009origin}, it is the ``heat of the compressed vacuum'' that accumulates along said symmetry line (i.e., as reservoir of ``outward'' oriented  kinetic energy) and therefore repels the trajectories. 
Fig.~\ref{fig:2} is in full concordance with the Bohmian interpretation (see, for example, \cite{Sanz.2008trajectory} for comparison). 
However, as mentioned, in our case also a ``micro-causal'' explanation is provided, which brings the whole process into perfect agreement with momentum conservation on a more ``microscopic'' level.

This can be shown even in greater detail.
Whereas in the example of Fig.~\ref{fig:2} the small amount of dispersion is practically negligible, we now discuss in more detail an interference scenario with significant dispersion of the two Gaussians, i.e., where initially the two Gaussians spread independently from each other due to the action of the diffusive velocities $\VEC{u}_1$ and $\VEC{u}_2$, respectively.
\begin{figure}[!t]
 \center
		\includegraphics[width=130mm,height=130mm]{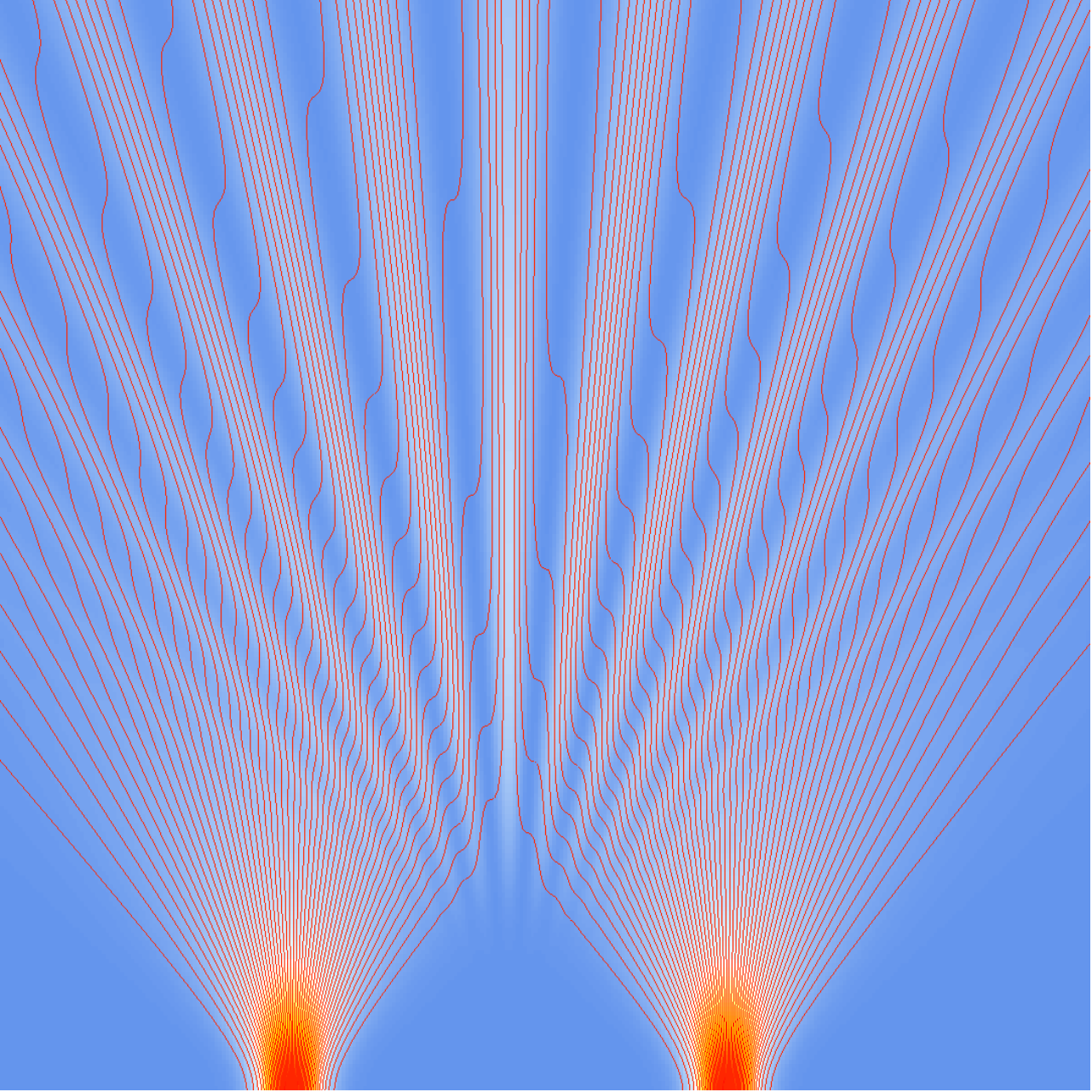}
		\caption{Classical computer simulation of the interference pattern: intensity distribution with increasing intensity from white through yellow and orange, with trajectories (red) for two Gaussian slits, and with large dispersion (evolution from bottom to top; $v_{x,1}=v_{x,2}=0$).
		The interference hyperbolas for the maxima characterize the regions where the phase difference $\varphi=2 n\pi$, and those with the minima lie at $\varphi=(2n+1)\pi$, $n=0,\pm1,\pm2,\ldots$
		Note in particular the ``kinks'' of trajectories moving from the center-oriented side of one relative maximum to cross over to join more central (relative) maxima.
		In our classical explanation of double slit interference, a detailed ``micro-causal'' account of the corresponding kinematics can be given.}
		\label{fig:3}
\end{figure}
In Fig.~\ref{fig:3}, trajectories according to Eqs.~\eqref{eq:3.12} and \eqref{eq:3.13} for the two Gaussian slits are shown.
The interference hyperbolas for the maxima characterize the regions where the phase difference $\varphi=2 n\pi$, and those with the minima lie at $\varphi=(2n+1)\pi$, $n=0,\pm1,\pm2,\ldots$
Note in particular the ``kinks'' of trajectories moving from the center-oriented side of one relative maximum to cross over to join more central (relative) maxima.
The trajectories are in full accordance with those obtained from the Bohmian approach, as can be seen by comparison with references~\cite{Holland.1993}, \cite{Bohm.1993undivided} and \cite{Sanz.2009context}, for example.
In our classical explanation of double slit interference, a detailed ``micro-causal'' account of the corresponding kinematics can be given:
Firstly, we note that the last term in Eq.~\eqref{eq:3.12}, which is responsible for the genuinely ``quantum'' behavior, is characterized by the vector
$\meanx{\VEC{u}}_1 - \meanx{\VEC{u}}_2$ which in the interference region is always oriented into the ``forward'' direction away from the slits (Fig.~\ref{fig:1}).
This means that said last term is modulated by $\sin\varphi$, with the result that the term alternates between ``forward'' directions where $\sin\varphi>0$ and ``backward'' directions where $\sin\varphi<0$.

Thus, in the ``backward'' cases, the trajectories coming from the relative maxima (bright fringes) loose velocity/momentum in the forward direction and cross over into the area of the relative minimum (dark fringes). Alternatively, in the ``forward'' cases, the trajectories coming from the relative minima (dark fringes) gain velocity/momentum in the forward direction and thus align with the other trajectories of the bright fringes.
In other words, for the areas where $\sin\varphi<0$, part of the current (along a relative maximum) is being removed (``depletion''), whereas for 
$\sin\varphi>0$, parts of currents flow together to produce a newly formed bright fringe (``accumulation'').
This is in accordance with our earlier description of quantum interference, where the effects of diffusion wave fields were explicitly described by alternating zones of heat accumulation or depletion, respectively \cite{Groessing.2009origin}.
Towards the central symmetry line, then, one observes heat accumulation from both sides, and due to big momentum kicks from the central accumulation of heat energy, the forward particle velocities' directions align parallel to the symmetry axis.
With the crossing-over of particle trajectories being governed by the last, diffusion-related, term on the right hand side of Eq.~\eqref{eq:3.12}, one finds that for $\varphi=0$ the resulting diffusive current is zero and thus, as total result of the overall kinematics, no crossing is possible.
Further, we note that our results are also in agreement with the recently published experimental results by Kocsis~et~al.~\cite{Kocsis.2011observing}.
Here we just comment that, as opposed to the Bohmian interpretation, we give a micro-causal explanation of these results without invoking the quantum potential.

\begin{figure}[!htp]
 \center
		\includegraphics[width=70mm,height=70mm]{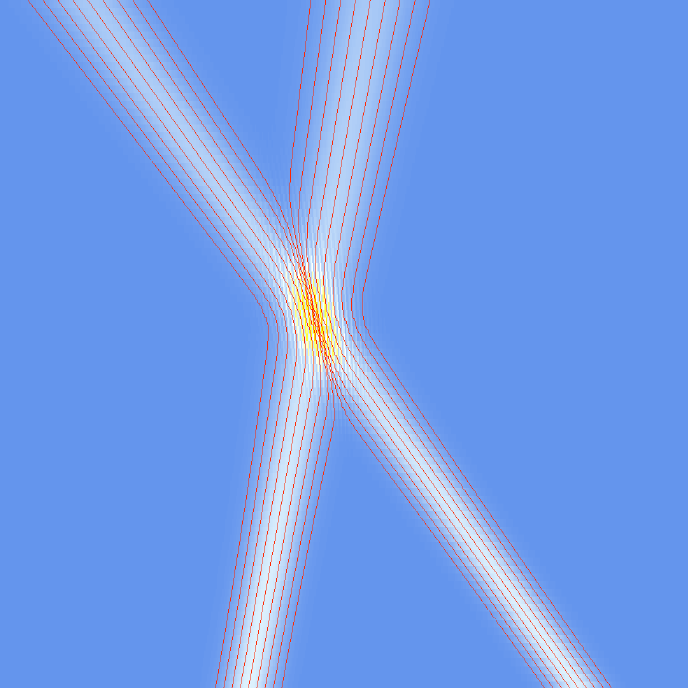}
		\caption{Same as Fig.~\ref{fig:2}, but with different initial average velocities: $v_{x,2}=-4v_{x,1}$.
		Note again the ``no crossing'' behavior, with the two trajectory bundles repelling each other due to the kinetic (heat) energy along the slanted central line. }
		\label{fig:2b}
\end{figure}

\begin{figure}[!htp]
 \center
		\includegraphics[width=70mm,height=70mm]{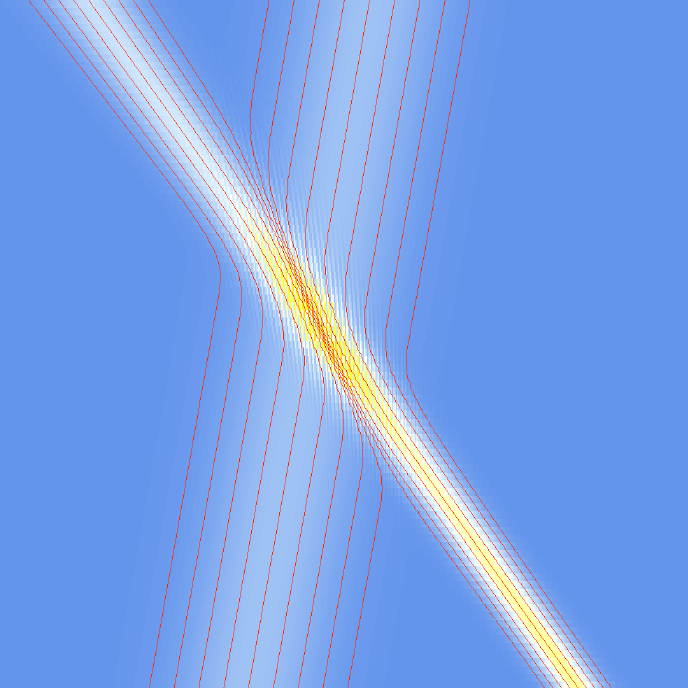}
		\caption{Same as Fig.~\ref{fig:2b}, but with different initial spreading: $\sigma_1=3\sigma_2$.
		Although the two partial beams altogether reflect off each other, one can clearly observe the effect of ``microscopic'' momentum conservation: the path excitation field of the right beam is propagated over to the micro-kinematics of the left beam, and \textit{vice versa}. }
		\label{fig:2c}
\end{figure}

\begin{figure}[!htp]
 \center
		\includegraphics[width=70mm,height=70mm]{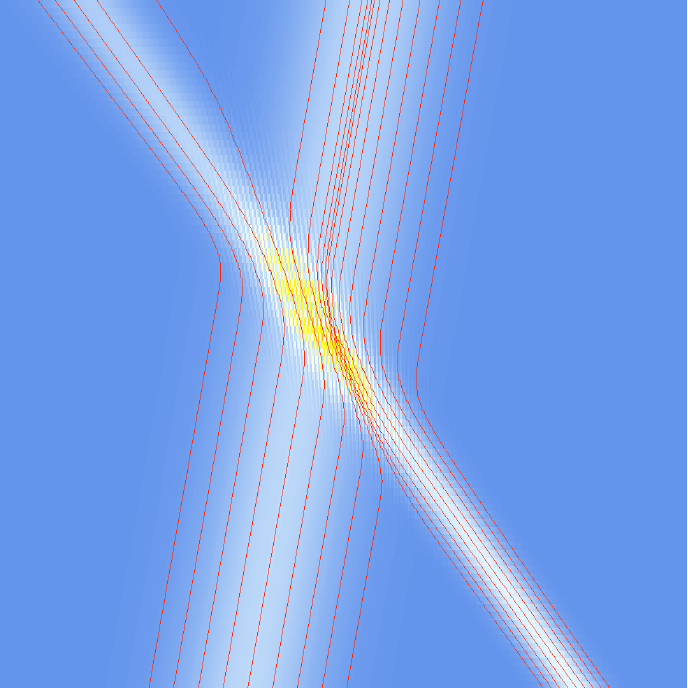}
		\caption{Same as Fig.~\ref{fig:2c}, but with different probability densities: $P_1=2P_2$. 
		Note that the emerging beam behavior compares more with inelastic scattering than with the elastic-type scattering of Fig.~\ref{fig:2c}, as part of the left beam merges with the right one. }
		\label{fig:2d}
\end{figure}

Finally, to illustrate the straightforward applicability of our model to more general situations, i.e., as compared to the simple symmetrical scenarios of Figs.~\ref{fig:2} and \ref{fig:3}, we now employ our simulation schema to cases where neither the Gaussians are identical nor their weights.
We thus study asymmetric coherent superpositions as discussed in reference~\cite{Sanz.2008trajectory}, and in our Figs.~\ref{fig:2b} to \ref{fig:2d} we show results in accordance with the figures 4-6 of ref.~\cite{Sanz.2008trajectory}.
The analysis of ref.~\cite{Sanz.2008trajectory} holds identically in our approach, so that we here restrict ourselves to pointing out that our figures display the following cases of varied properties for the beams emerging from the two slits:
\begin{itemize}
	\item different modulus of the initial velocity/momentum,
	\item different initial spreading,
	\item different weights for the probability densities.
\end{itemize}

\section{The Schr\"odinger equation revisited}
\label{sec:schrodinger}

Upon employment of the Madelung transformation for each path $j$ ($j=1$ or $2$),
\begin{equation}  \label{eq:3.14}
  \Psi_j = R\e^{iS_j/\hbar} ,
\end{equation}
and thus $P_j=R_j^2=|\Psi_j|^2=\Psi_j^*\Psi_j$, with the definitions \eqref{eq:2.7} and $v_j:=\nabla S_j/m$, $\varphi=(S_1-S_2)/\hbar$, and recalling the usual trigonometric identities such as
$\cos\varphi=\frac{1}{2}\left(\e^{i\varphi}+\e^{-i\varphi}\right)$, one can rewrite the total average current \eqref{eq:3.12} immediately as
\begin{equation}  \label{eq:3.18}
  \begin{array}{rl}
    \displaystyle
      \meanx{\VEC{J}}_{\rm tot} &= P_{\rm tot}\meanx{\VEC{v}}_{\rm tot} \\ [3ex]
          &= \displaystyle(\Psi_1+\Psi_2)^*(\Psi_1+\Psi_2) \frac{1}{2}\left[ 
              \frac{1}{m}\left(-i\hbar \frac{\nabla(\Psi_1+\Psi_2)}{(\Psi_1+\Psi_2)} \right)+
              \frac{1}{m}\left( i\hbar \frac{\nabla(\Psi_1+\Psi_2)^*}{(\Psi_1+\Psi_2)^*} \right)\right] \\ [3ex]
          &= \displaystyle-\frac{i\hbar}{2m}\left[ \Psi^*\nabla\Psi - \Psi\nabla\Psi^*\right]
					= \displaystyle\frac{1}{m}{\rm Re}\left\{ \Psi^*(-i\hbar\nabla)\Psi \right\}, 
  \end{array} 
\end{equation}
where $P_{\rm tot}=|\Psi_1+\Psi_2|^2=:|\Psi|^2$.
In the Appendix we show an alternative derivation of Eq.~\eqref{eq:3.18} from our classical approach.
The last two expressions of \eqref{eq:3.18} are the exact well-known formulations of the quantum mechanical probability current, here obtained without any quantum mechanics, but just by a re-formulation of \eqref{eq:3.12}. In fact, it is a simple exercise to insert the wave functions \eqref{eq:3.14} into \eqref{eq:3.18} to reobtain \eqref{eq:3.12}.

Moreover, one obtains with \eqref{eq:3.18} for the total average current the conservation law
\begin{equation}  \label{eq:3.19}
  \frac{\p P_{\rm tot}}{\p t} = -\nabla\cdot\meanx{\VEC{J}}_{\rm tot} = \frac{i\hbar}{2m} \nabla\left[ \Psi^*\nabla\Psi - \Psi\nabla\Psi^*\right] .
\end{equation}
As it also holds that
\begin{equation}  \label{eq:3.20}
  \frac{\p P_{\rm tot}}{\p t} = \frac{\p}{\p t}\left(\Psi^*\Psi\right) = \Psi^*\dot{\Psi} + \Psi\dot{\Psi}^*  ,
\end{equation}
a comparison of Eq.~\eqref{eq:3.19} and \eqref{eq:3.20} provides
\begin{equation}  \label{eq:3.21}
  i\hbar \left[\Psi^*\dot{\Psi} + \Psi\dot{\Psi}^*\right] = -\frac{\hbar^2}{2m}\left[ \Psi^*\nabla^2\Psi - \Psi\nabla^2\Psi^*\right] ,
\end{equation}
which is the difference between two equations, i.e.,
\begin{equation}  \label{eq:3.22}
  i\hbar \dot{\Psi}\Psi^* = \left(-\frac{\hbar^2}{2m}\nabla^2\Psi + V\Psi\right)\Psi^*
\end{equation}
and
\begin{equation}  \label{eq:3.23}
  -i\hbar \dot{\Psi}^*\Psi = \left(-\frac{\hbar^2}{2m}\nabla^2\Psi^* + V\Psi^*\right)\Psi
\end{equation}
and thus of the Schr\"odinger equation (with external potential $V$)
\begin{equation}  \label{eq:3.24}
  i\hbar \dot{\Psi} = -\frac{\hbar^2}{2m}\nabla^2\Psi + V\Psi
\end{equation}
and its charge conjugate.

The validity of the Schr\"odinger equation is here seen to be based on the assumption constitutive for our model, i.e., of the quantum being an off-equilibrium steady-state system, with its oscillations $\omega$ being maintained by a steady and balanced throughput of energy. 
This raises, of course, the possibility to consider the breakdown of that balance, which is conceivable at time scales $\tau\simeq 1/\omega$ or shorter. 
For, we have assumed throughout that said balance is maintained between diffusive currents, which are responsible for the maintenance of the oscillation of frequency $\omega$. 
So, an eventual breakdown of the Schr\"odinger equation and the usual laws of quantum mechanics can be assumed when the symmetry between diffusive currents is broken, i.e., on the mentioned timescales.

\section{Conclusions and outlook}
\label{sec:conclusion}

Considering ``particles'' as oscillators (``bouncers'') coupling to regular oscillations of the ``vacuum's'' zero-point field, which they also generate, we have shown how a quantum can be understood as an emergent system. In particular, the dynamics
between the oscillator and the ``bath'' of its thermal environment can be made responsible not only for Gaussian diffraction at a single slit \cite{Groessing.2010emergence}, but also for the well-known interference effects at a double slit. 
We have also shown how the model entails the existence of a path excitation field, i.e., a field spanned by the average velocity fields 
$\meanx{\VEC{v}}(\VEC{x},t)$
and $\meanx{\VEC{u}}(\VEC{x},t)$, respectively, where the latter refer to diffusion processes reflecting also the stochastic parts of the zero-point field. We have derived in this paper, on the basis of classical physics, i.e., without the use of quantum mechanical terms, the exact intensity distribution at a screen behind a double slit, as well as the details of the more complicated particle current, or of the Bohmian particle trajectories, respectively.

In a computer simulation, moreover, we have pictured quantum interference employing well-known techniques to model classical diffusion processes. 
With this new tool at hand, and having proven the identity with the results of the usual quantum mechanical formalism, we are encouraged to tackle further foundational issues, such as relativistic formulations, and, in particular, questions concerning nonlocality and entanglement. 
As to the latter, one thing can be noted right away. Considering double interferometry, for example, with the emission of two anti-correlated particles with opposite directions, the nature of the path excitation field immediately implies two features: a) the ``wholeness'' of the processes going on in the total domain of the experimental setup, and b) nonlocal correlations between distant parts of the experiment. 

For, it is obvious that, say, the very existence of the initial Gaussian representing the source (and at some time ``decaying'' into two Gaussians propagating in opposite directions), implies one path excitation field in \textit{both} directions. 
This means that whereas the average velocities for the respective particles, $\meanx{\VEC{v}}$ and $\meanx{\VEC{v}}'$ are anti-parallel and refer to separate regions in space, it is one and  the same diffusive velocity field $\meanx{\VEC{u}}\equiv\meanx{\VEC{u}}'$ that affects both particles throughout the experimental setup! 
We shall show in a forthcoming paper how this feature of ``wholeness'' implies the existence of nonlocal correlations. 
Due to the nonlocal nature of the involved diffusion wave fields, and based on our proposed model, it should be possible to prove a corresponding identity with entangled states in quantum mechanics.

The phenomenon of entanglement is thus possibly rooted in the existence of the path excitation field, a version of which we already encountered in the present paper, i.e., with the usual double slit interferometry. 
So, one can expect that the concepts developed here will also help to deepen our understanding of entanglement. 
As far as explanations of the physics of the double slit experiment are concerned, however, we think that we can safely say that it is possible to understand quantum mechanics.

\appendix

\section{}
\label{sec:appendix}

In this appendix, we give an alternative derivation of the quantum mechanical current \eqref{eq:3.18} from classical physics.
We have in previous papers employed the Madelung transformation
\begin{equation}  \label{eq:6.1}
  \Psi = R\e^{iS/\hbar} ,
\end{equation}
and thus $P=R^2=|\Psi|^2=\Psi^*\Psi$, to obtain a ``vocabulary'' for the translation of our notation into that of quantum mechanics,
\begin{equation}  \label{eq:6.2}
  \meanx{p^2}_{\rm tot} = \hbar^2\meanx{k^2}_{\rm tot} = 
      \hbar^2\meanx{\left|\frac{\nabla \Psi}{\Psi} \right|^2} = \hbar^2 \int P \left|\frac{\nabla \Psi}{\Psi} \right|^2 \d^n x =
        \meanx{\left(\frac{\nabla S}{\hbar} \right)^2}
      + \meanx{\left(\hbar\frac{\nabla R}{R} \right)^2}
      = m^2\meanx{v^2} + m^2\meanx{u^2},
\end{equation}
which provides also that \cite{Groessing.2010entropy} 
\begin{equation}  \label{eq:6.3}
  P_{\rm tot} = \meanx{|\Psi_{\rm tot}|^2} = \frac{\meanx{|\nabla\Psi_{\rm tot}|^2}}{\meanx{k_{\rm tot}^2}}.
\end{equation}
(Note that generally the division by $\Psi$ is regular, since wherever $\VEC{u}$ is defined, a particle path exists, and therefore $\Psi \neq 0$.)
Thus, one can also write down the momentum vector $\meanx{\VEC{p}}_{\rm tot}$ as a complex one providing
\begin{equation}  \label{eq:6.4}
  \meanx{\VEC{p}}_{\rm tot}^{(+)} = m\meanx{\VEC{v}}_{\rm tot}^{(+)} = -i\hbar \frac{\nabla\Psi_{\rm tot}}{\Psi_{\rm tot}} \qquad\text{and}\qquad
	\meanx{\VEC{p}}_{\rm tot}^{(-)} = m\meanx{\VEC{v}}_{\rm tot}^{(-)} =  i\hbar \frac{\nabla\Psi_{\rm tot}^*}{\Psi_{\rm tot}^*}.
\end{equation}
(In quantum mechanical terms, Eq.~\eqref{eq:6.4} is the corollary of the fact that $\meanx{\VEC{p}}_{\rm tot}$ is the ``eigenvector'' of the ``momentum operator'' $\VEC{\hat{p}}=-i\hbar\nabla$.)
Moreover, as was shown in \cite{Groessing.2010entropy}, one can use the classical wave mechanics formula \eqref{eq:3.0}, 
$R\VEC{k}_{\rm tot} = R_1\VEC{k}_{{\rm tot},1} + R_2\VEC{k}_{{\rm tot},2}$, and Eq.~\eqref{eq:6.2} for the case of two alternative paths 1 and 2, to write 
$\meanx{k^2}_{\rm tot} = \meanx{\left|\frac{\nabla \Psi_{{\rm tot}}}{\Psi_{{\rm tot}}} \right|^2} = \int P\VEC{k}_{{\rm tot}}\cdot\VEC{k}_{{\rm tot}} \d^n x=\int |R_1\VEC{k}_{{\rm tot},1}+R_2\VEC{k}_{{\rm tot},2}|^2\d^nx$, which implies \eqref{eq:3.0a} and
\begin{equation}  \label{eq:6.5}
  P_{\rm tot} = \left|\Psi_{\rm tot}\right|^2 = \left|\Psi_1 + \Psi_2\right|^2 \qquad\text{corresponding to}\qquad
	\Psi_{\rm tot} = \Psi_1 + \Psi_2.
\end{equation}
If, with these ingredients, one now defines the average velocity 
$\meanx{\VEC{v}}_{\rm tot}:=\frac{1}{2}\left[\meanx{\VEC{v}}_{\rm tot}^{(+)} + \meanx{\VEC{v}}_{\rm tot}^{(-)}\right]$, one immediately has
\begin{equation}  \label{eq:6.6}
	\meanx{\VEC{v}}_{\rm tot} = 
		-\frac{i\hbar}{2m} \left[ \frac{\nabla(\Psi_1+\Psi_2)}{\Psi_1+\Psi_2} - \frac{\nabla(\Psi_1+\Psi_2)^*}{(\Psi_1+\Psi_2)^*} \right],
\end{equation}
from which one reobtains the total quantum mechanical current \eqref{eq:3.18}.

\end{document}